\documentclass[fleqn,usenatbib,letters]{mnras}

\usepackage{newtxtext,newtxmath}

\usepackage[T1]{fontenc}

\DeclareRobustCommand{\VAN}[3]{#2}
\let\VANthebibliography\thebibliography
\def\thebibliography{\DeclareRobustCommand{\VAN}[3]{##3}\VANthebibliography}

\usepackage{graphicx}	
\usepackage{amsmath}	
\usepackage[normalem]{ulem}

\newcommand{\psj}{PSJ}

\title[JWST spectrum of C/2024 E1 (Wierzchos)]{First JWST spectrum of distant activity in Long Period Comet C/2024~E1~(Wierzchos)}

\author[C. Snodgrass et al.]{
Colin Snodgrass,$^{1}$\thanks{E-mail: csn@roe.ac.uk (CS)}
Carrie E. Holt,$^{2}$
Michael S. P. Kelley,$^{3}$
Cyrielle Opitom,$^{1}$
Aurélie Guilbert-Lepoutre,$^4$
\newauthor
Matthew M. Knight,$^5$
Rosita Kokotanekova,$^{6,7}$
Emmanuel Jehin,$^8$
Elena Mazzotta Epifani,$^9$
\newauthor
Alessandra Migliorini,$^{10}$
Cecilia Tubiana,$^{10}$
Marco Micheli,$^{11}$
Davide Farnocchia$^{12}$
\\
$^{1}$Institute for Astronomy, University of Edinburgh, Royal Observatory, Edinburgh EH9 3HJ, UK\\
$^{2}$Las Cumbres Observatory, 6740 Cortona Drive Suite 102, Goleta, CA 93117, USA\\
$^{3}$Department of Astronomy, University of Maryland, College Park, MD 20742, USA\\
$^4$ LGL-TPE, UMR 5276 CNRS, Université Lyon 1, ENS, Villeurbanne, France\\
$^5$ Physics Department, United States Naval Academy, 572C Holloway Road, Annapolis, MD 21402, USA\\
$^6$ Institute of Astronomy and National Astronomical Observatory, Bulgarian Academy of Sciences, 72 Tsarigradsko Shose Boulevard, 1784 Sofia, Bulgaria\\
$^7$ International Space Science Institute, Hallerstrasse 6, 3012 Bern, Switzerland\\
$^8$ STAR Institute, University of Liège, Allée du 6 août, 19, 4000 Liège (Sart-Tilman), Belgium\\
$^9$ INAF - Osservatorio Astronomico di Roma, Via Frascati 33, Monte Porzio Catone (RM), Italy\\
$^{10}$ Institute for Space Astrophysics and Planetology, IAPS-INAF, Rome, Italy\\
$^{11}$ ESA NEO Coordination Centre, Largo Galileo Galilei, 1, 00044 Frascati (RM), Italy\\
$^{12}$ Jet Propulsion Laboratory, California Institute of Technology, 4800 Oak Grove Dr., Pasadena, 91109, CA, United States
}

\date{Accepted XXX. Received YYY; in original form ZZZ}

\pubyear{\the\year{}}

\begin{document}
\label{firstpage}
\pagerange{\pageref{firstpage}--\pageref{lastpage}}
\maketitle

\begin{abstract}
We observed the new Long Period Comet C/2024 E1 (Wierzchos), inbound at 7 au from the Sun, using the NIRSpec integral field unit on JWST. The spectrum shows absorption features due to water ice in the coma and evidence for CO$_2$ driven activity, with a production rate of $Q(CO_2) = 2.546 \pm 0.019 \times 10^{25}$ molecules s$^{-1}$, and no emission features of water or CO. The latter is surprising, given that CO is more volatile than CO$_2$, and suggests that this comet may have lost its near-surface CO during its early evolution, before implantation in the Oort cloud.
\end{abstract}




\section{Introduction}

Historically, Long Period Comets (LPCs) have been seen as short-lived phenomena -- temporary visitors to our skies observed for a few months only, as they pass close to the Sun and put on a show of activity that is primarily driven by sublimation of water ice, before returning to the outer reaches of our Solar System and being lost from view. Over the past decade or so, modern sky surveys increasingly discover LPCs at larger distances from the Sun \citep{Lister-LOOK-summary}. Activity is now regularly observed at heliocentric distance $r_h > 5$ au \citep[e.g.][]{MazzottaEpifani2009,Meech2009,Rousselot2014,Holt2024}. At these distances temperatures are low enough that water ice does not sublimate efficiently and more volatile species, such as CO or CO$_2$, are likely activity drivers \citep[e.g.][]{Meech-CometsII}. A handful of exceptional comets have been observed to be active at distances beyond 20 au from the Sun, such as C/2010 U3 (Boattini), which was active at 25.8 au \citep{Hui2019}, C/2017 K2 (Pan-STARRS) at 23.7 au  \citep{Hui2018,Jewitt2021}, C/2014 UN271 (Bernardinelli-Bernstein) at 26 au \citep{Farnham2021}, and C/2019 E3 (ATLAS) at 23 au \citep{Hui2024}. Discovery at such large distances allows study of their evolving activity drivers with changing $r_h$, although direct observation of the expected activity drivers for comets (H$_2$O, CO, or CO$_2$) remains challenging, as the Earth's atmosphere prevents observations of these species' infrared emission features from the ground.

Discovery of LPCs at large distance also greatly increases the time over which these comets can potentially be observed, from months to years, and opens the possibility of sending a space mission to encounter one. This is what the European Space Agency (ESA)'s {\it Comet Interceptor} mission (CI) will do in the first half of the next decade; the expected increase in discovery rate of LPCs at large distance with the beginning of the Vera C. Rubin observatory's Legacy Survey of Space and Time (LSST - \citealt{LSST}) will mean that a spacecraft waiting in space could be directed towards a LPC flyby with a few years warning time \citep{Snodgrass+Jones-CI, Jones-etal-CI}. However, not all LPCs are reachable by this mission; due to spacecraft constraints it will need to encounter its comet close to $r_h = 1$ au, and many of the LPCs discovered at large distances do not get that close to the Sun -- C/2014 UN271, for example, will approach no closer than the orbit of Saturn. A suitable mission target will be one discovered at large $r_h$ but with perihelion inside of $r_h = 1.2$ au (among other constraints -- \citealt{Pau-CI}).

C/2024 E1 (Wierzchos), hereafter E1, was discovered by the Catalina sky survey in March 2024 at $r_h = 8$ au \citep{MPEC} and will have its perihelion passage at 0.6 au in January 2026. While this will be before CI will be operating, this comet presents an excellent opportunity to study how the activity drivers of such comets evolve as they approach the Sun. Understanding the behavior of comets similar to the expected CI target is of high priority for mission success. From a mission planning point of view, it is essential to be able to predict the activity level and expected production rates close to perihelion based on distant observations. On the other hand, having a detailed understanding of the range of properties and activity patterns observed among LPCs is necessary to put CI's measurements of an individual object into the larger context of the population. E1 was the first comet with such an orbit to be discovered at such extreme distance during the JWST-era, and since ESA selected the CI mission in 2019, giving us an excellent opportunity to learn about cometary evolution over relevant distances before the mission launches.

JWST promises to be a revolutionary facility in the study of comets \citep{Kelley2016}. As a space telescope with a wavelength range (for NIRSpec) covering the 1--5 micron infrared range it is sensitive to emission features of all major drivers of cometary activity (H$_2$O, CO, CO$_2$) and many of the other minor species that are expected to sublimate directly from ices (as opposed to the gas species observed in the UV/visible region, that are mostly the products of photo-dissociation in the coma; e.g., \citealt{Bodewits-CometsIII}). The NIRSpec integral-field-unit (IFU) mode allows not only the detection of these species, but mapping of their spatial distribution, which has revealed heterogeneous outgassing in the coma of comet 29P/Schwassmann-Wachmann~1 \citep{Faggi2024}.
JWST's sensitivity is opening a new window on distant or weak cometary activity, and it has already provided the first direct evidence for water ice sublimation driven activity in Main Belt Comets \citep{Kelley2023,Hsieh2025} and CO$_2$  and methane emissions in Centaurs \citep{HarringtonPinto2023,Pinilla-Alonso2024-Chiron}. 

In this letter we present the first JWST spectrum of a LPC beyond the water ice sublimation region. These results are from the first of three epochs approved in our JWST programme, obtained when the comet was at 7 au from the Sun; we will observe E1 two more times in 2025, covering the full range of its pre-perihelion visibility from JWST at approximately 5 and 3 au, and, in a future work, will study how its spectrum and activity pattern evolve as the comet crosses the water ice line.

\section{Observations and Data Reduction}

We observed the comet using JWST/NIRSpec \citep{NIRSpec} in IFU prism mode, covering the full wavelength range  0.6 -- 5.3 \micron{} at low resolution ($R \approx 30$ to 300, variable with wavelength). Observations were performed on 2024 June 14th around 21:00 UT, when the comet was at 7.2 au from the Sun, with a distance of 7.0 au and a phase angle of 8.2\degr{} from JWST. At this time ground-based observations by our team using the Telescopio Nazionale Galileo, Las Cumbres Observatory Global Telescope network, and TRAPPIST telescopes  showed a faint and condensed coma with a total $r$-band magnitude of 19.7, Solar-like colour, and $Af\rho \approx 300$ cm. Details of ground-based observations of E1, including longer-term monitoring and multi-filter photometry, will be published in future papers. For our JWST/NIRSpec observations we used the NRSIRS2RAPID readout mode with 50 groups per integration, 2 integrations per exposure, and a 4-point dither pattern that kept the comet within the 3\arcsec{} field-of-view of the IFU, resulting in a total exposure time of 5952 s. We also obtained an offset sky observation 5\arcmin{} from the comet, as the comet was expected to fill the IFU field-of-view.

The uncalibrated data were downloaded from the Mikulski Archive for Space Telescopes and processed using version 1.16.1 of the JWST Science Calibration Pipeline with JWST Calibration Reference Data System context file 1321. The background sky exposures were subtracted from the exposures of the comet. The four comet exposures were then combined by resampling the dithered observations to create a 53 $\times$ 51 $\times$ 941 pixel data cube. A comet spectrum was extracted using a 0.4\arcsec{} ($\sim$2000 km) radius aperture centred on the brightest point (assumed to coincide with the nucleus).

\section{Results}

\begin{figure}
	\includegraphics[width=\columnwidth]{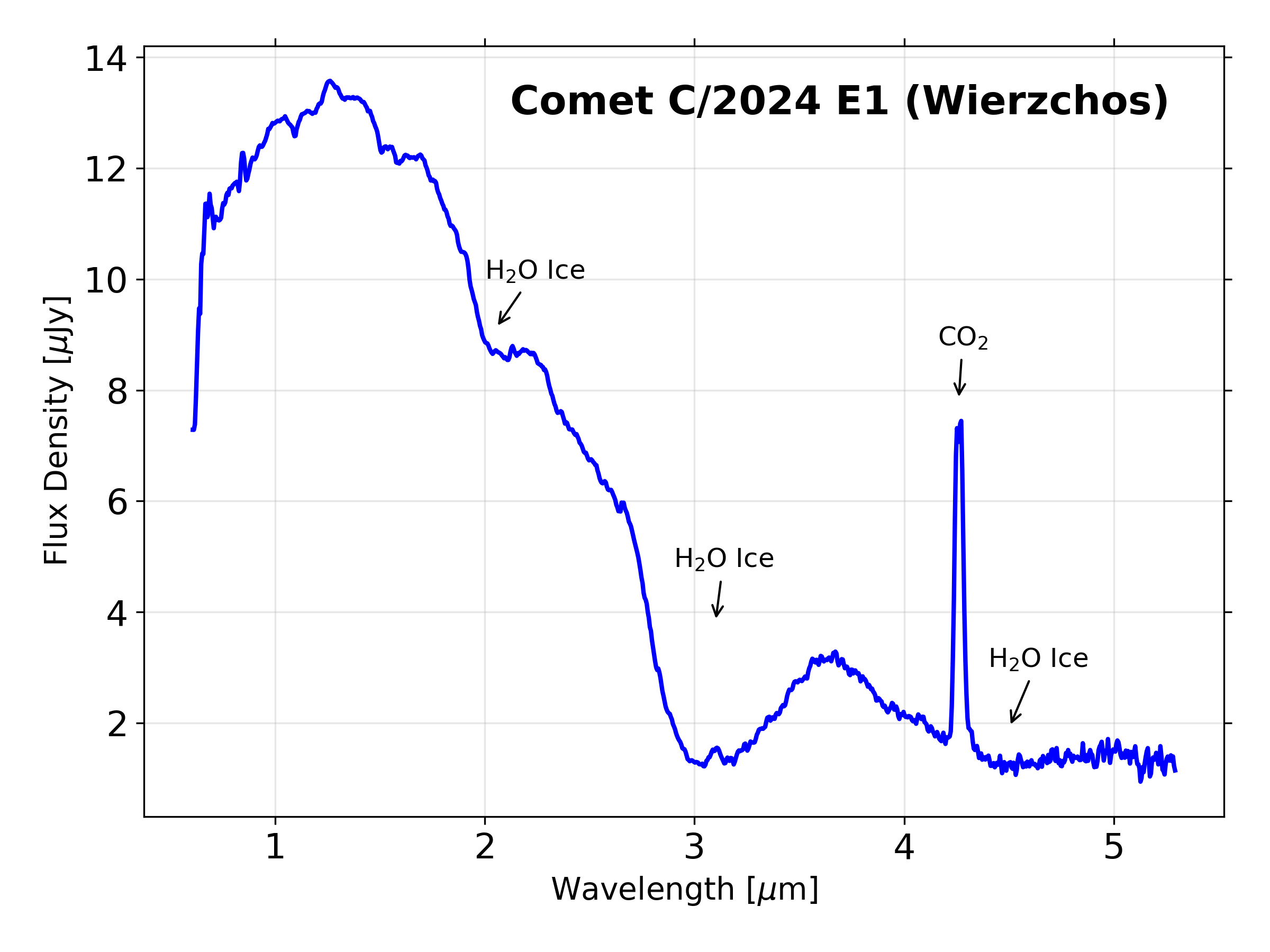}
    \caption{Spectrum extracted within a 0.4\arcsec{} radius aperture centered on the nucleus.}
    \label{fig:spectrum}
\end{figure}

\begin{figure}
	\includegraphics[width=\columnwidth]{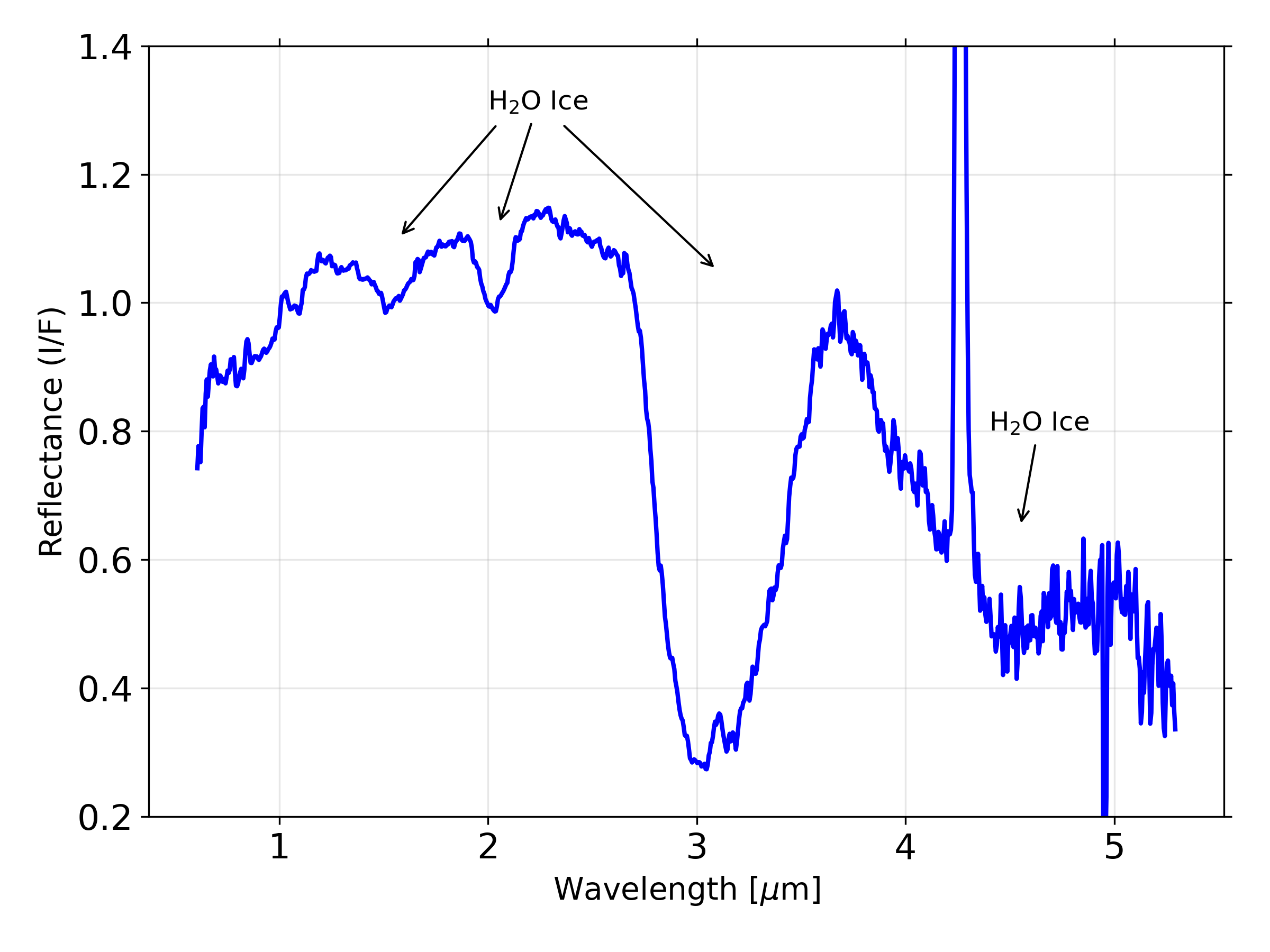}
    \caption{Reflectance spectrum within the same photometric aperture as in Fig~\ref{fig:spectrum}}
    \label{fig:reflectance}
\end{figure}

\subsection{Spectrum}

In Fig.~\ref{fig:spectrum} we show the extracted spectrum of the comet. The  most prominent features are labelled: these are a broad absorption around 3 \micron{} attributed to water ice grains, and a strong emission from CO$_2$ at 4.3 \micron. Emission from water (2.69~\micron) or CO gas (4.67~\micron) is notably absent. The appearance of the Fresnel peak in the centre of the water ice band at around 3.1 \micron{} indicates that the ice is crystalline \citep[e.g.,][]{Mastrapa2009}. There are also weaker absorption features visible at 1.5, 2 and 4.5 \micron{} that can be attributed to  water ice \citep{Mastrapa2013}, which are more easily seen in Fig.~\ref{fig:reflectance}, which shows the reflectance spectrum obtained by dividing Fig.~\ref{fig:spectrum} by the spectrum of the G-type solar analog star SNAP-2 (Program \#1128; PI: N. Luetzgendorf; publicly available data). 

\begin{figure*}
	\includegraphics[width=\textwidth]{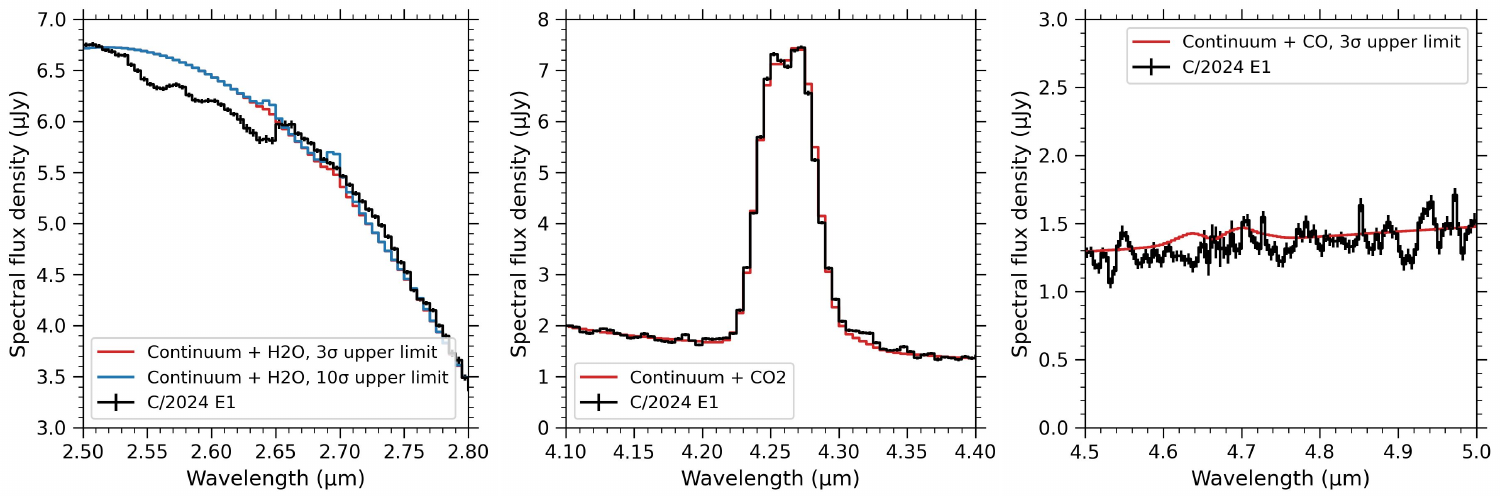}
    \caption{Zoom in on the expected gas emission bands, showing the models used to derive the CO$_2$ production rate and upper limits for H$_2$O and CO.}
    \label{fig:spectrum-fit}
\end{figure*}

In order to quantify the CO$_2$ production rate, we performed a retrieval of the extracted spectrum (Fig. \ref{fig:spectrum}) using the Planetary Spectrum Generator \citep[PSG;][]{Villanueva-PSG}. We focused the retrieval on the CO$_2$ $\nu_3$ band at 4.26 \micron{} by subtracting a continuum modeled using a polynomial fit and limiting the spectral range to 4.1 - 4.4 \micron{}. The best fit from the PSG resulted in a derived production rate of $Q(CO_2) = 2.546 \pm 0.019 \times 10^{25}$ molecules s$^{-1}$. We fit coma temperature as a free parameter and a rotational temperature of $59.0 \pm 0.9$ K was derived to match the shape of the emission band. The best-fit fluorescence model is shown in Fig.\ref{fig:spectrum-fit}.

Fluorescence emission from the H$_2$O $\nu_3$ band at 2.69~\micron{} and the CO 1$\rightarrow$0 band at 4.67~\micron{} may be used to estimate upper-limit production rates for these species.  The PSG was once again used to generate model spectra, using the same coma expansion speed as for the CO$_2$ coma.  To fit the data, additional considerations were taken to account for correlated noise in the IFU data, similar to the approaches previously used with NIRSpec Prism observations of comets 238P and 358P \citep{Kelley2023, Hsieh2025}.  To that end, a three component model was used, including: (1) the spectrum of the gas being tested: (2) a polynomial model for the continuum (first-order for CO, second-order for H$_2$O); and (3) a two-parameter model to describe the correlation between data points.  Gaussian process regression and a Markov Chain Monte Carlo technique are used to model the spectrum and parameter uncertainties.

An assumption on the rotational temperature is also needed to derive an upper-limit.  The CO$_2$ molecule lacks a dipole moment \citep[0.000 debye;][]{Nelson_1967}, and its rotational temperature is expected to remain constant with increasing distance outside of the molecular collisional zone near the nucleus ($\ll100$~km).  CO has a weak dipole moment \citep[0.112 debye;][]{Nelson_1967}, so we assume the same rotational temperature as measured for CO$_2$.  H$_2$O has a significantly stronger dipole \citep[1.857 debye;][]{Shostak1991}, therefore we assume the PSG's default temperature at 7.2~au: 8.4~K.  Given these assumptions, the best-fit 99.7\% confidence limits (equivalent to 3$\sigma$) are $2.1\times10^{24}$ and $8.1\times10^{24}$~molecules~s$^{-1}$, for H$_2$O and CO, respectively (Fig. \ref{fig:spectrum-fit}).

Our measurement of the CO$_2$ production rate and limits on the other major species allow us to put limits on abundance ratios. The limit on CO$_2$/H$_2$O $>$ 12 is in line with the general trend for other comets \citep{HarringtonPinto2022} as water ice sublimation becomes inefficient with increasing heliocentric distance \citep{Meech-CometsII}. The CO/CO$_2$ $<$ 32\% ratio is not unusual, although still an outlier in the survey by \citet{HarringtonPinto2022}, who found a general trend of increasing CO/CO$_2$ with heliocentric distance, and observed only CO-dominated comets beyond 3.5 au, based on observations of four comets.

\begin{figure*}
    \centering
    \includegraphics[width=\textwidth]{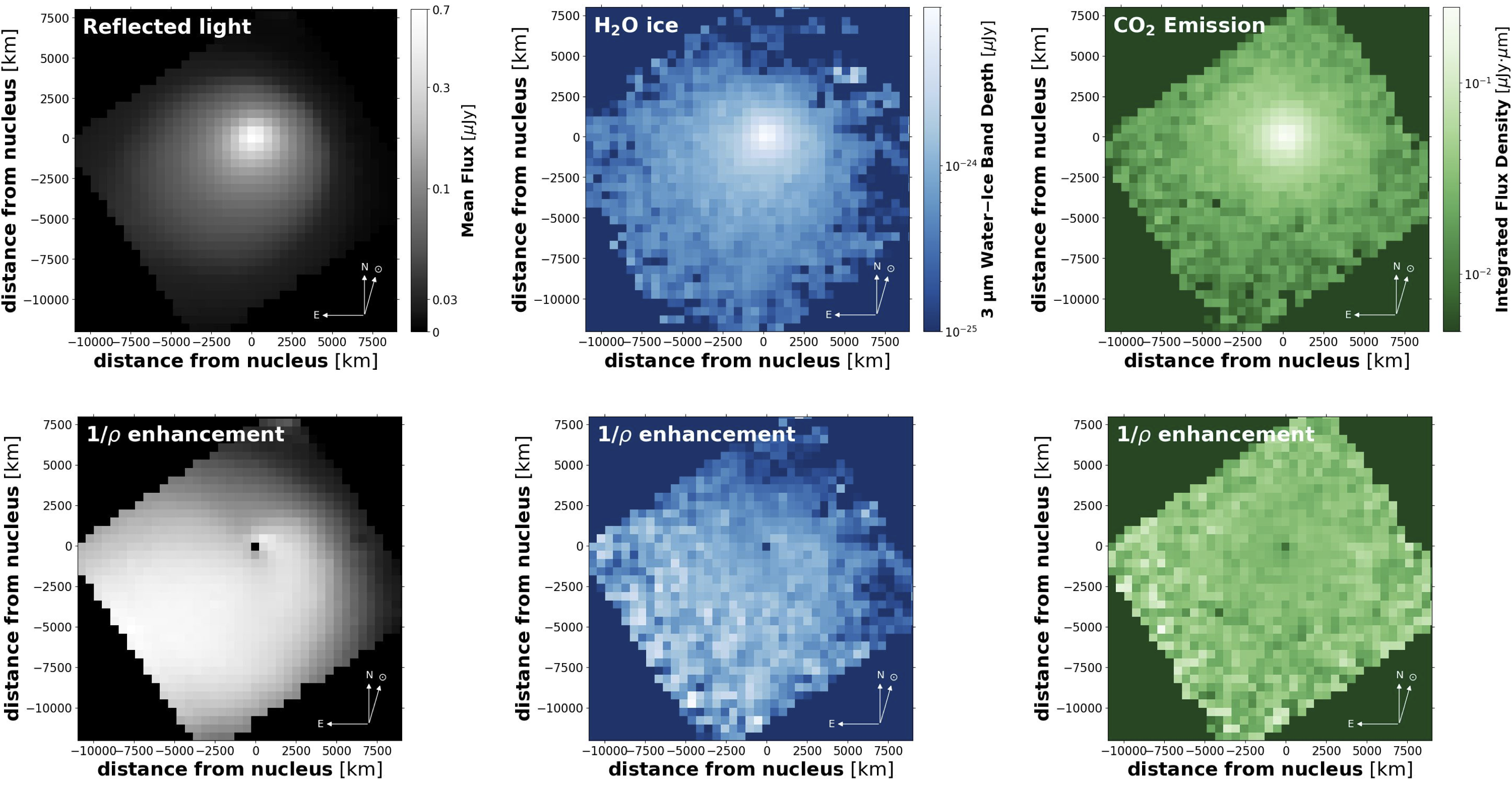}
    \caption{Images of the comet created from IFU-cubes that cover reflected continuum (i.e., dust) over 0.7 - 2.5 \micron{} (left), water ice absorption depth (3 \micron; centre), and CO$_2$ emission (4.22 - 4.31 \micron; right). The lower row shows the same images enhanced by dividing by a 1/$\rho$ profile to reveal asymmetries.}
    \label{fig:maps}
\end{figure*}

\subsection{Spatial distribution}

The observations were designed to use NIRSpec IFU mode so that we could also investigate the spatial distribution of the dust, ice, and gas in the comet's coma. Fig.~\ref{fig:maps} shows maps derived using different methods for each spectral feature. The dust image was constructed using the average flux value of the dust continuum in the wavelength range 0.7 - 2.5 \micron{}. The water ice absorption map was generated by calculating the depth of the 3.0 \micron{} absorption band, defined as the difference between the average band flux from 2.95 to 3.05 \micron{} and a linearly interpolated continuum after dividing the spectrum by the same solar analog spectrum used to create Fig.~\ref{fig:reflectance}. The CO$_2$ emission map was produced by integrating the continuum-subtracted flux over the 4.22 – 4.31 \micron{} wavelength range. 

\begin{figure}
	\includegraphics[width=\columnwidth]{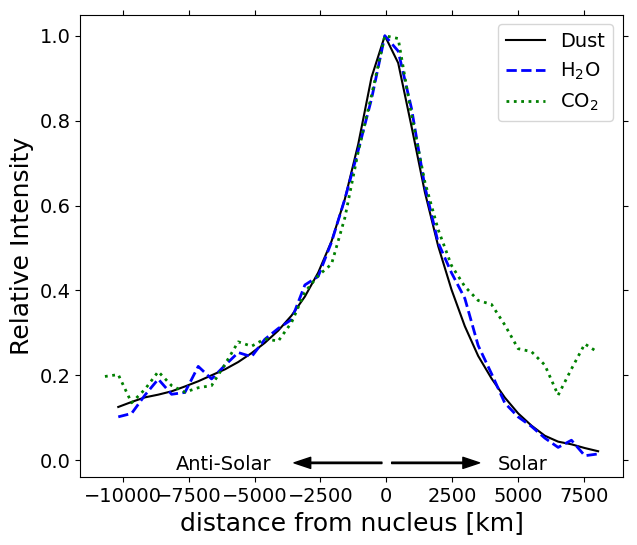}
    \caption{The solar/anti-solar radial profile of the coma from the three spectral regions shown in Fig.~\ref{fig:maps}}
    \label{fig:profile}
\end{figure}

The dust and ice images both show an asymmetric coma extending in the anti-solar direction, and have a similar distribution, indicating that the ice detected consists of icy grains in the coma. The CO$_2$ emission has a much more symmetric distribution. The lower panels in Fig.~\ref{fig:maps} show the images enhanced by division by a 1/$\rho$ radial profile (i.e., the theoretical profile of a freely expanding coma), which helps identify spatial asymmetries in the coma. The gas profile is highly symmetric and distinct from the dust and ice.  The latter show some evidence for a spiral pattern with material leaving towards the west and curving clockwise towards the tail. The similarity between the dust and water ice distributions is also seen in Fig.~\ref{fig:profile}, which shows radial profiles along the Sun--anti-Sun line for the three maps. The CO$_2$ gas coma is also seen to be more symmetrical either side of the nucleus in this plot.

We extracted a J-band image by integrating over the spectral dimension after weighting the data cube by the J-bandpass curve (1.15--1.33 \micron{}). We then measured the flux in this image within a 0.26\arcsec{} (1.5$\times$ the PSF FWHM) radius aperture centered on the nucleus, to obtain a limit of the nucleus size from the possible contribution of a point source to the inner coma. We find an upper limit of 13.7 km for the nucleus radius, assuming a typical cometary linear phase function with a slope of 0.047 magnitudes per degree and 4\% albedo in the visible \citep{Knight-CometsIII}, which corresponds to
a 6\% albedo in the J-band assuming a 10\%/100 nm spectral slope between the visible and near infra-red.

\section{Discussion and Conclusions}

Considering the gas coma, the lack of water emission at 7 au was expected, but the apparent lack of CO in this comet is surprising. Previous surveys of CO vs CO$_2$ activity have found that distant comets (beyond 3.5 au) have CO-dominated activity \citep{HarringtonPinto2022}. JWST observations of 29P revealed both CO and CO$_2$ at 6 au from the Sun, with a global production rate ratio of CO/CO$_2$ = 70 \citep{Faggi2024}. 29P is a Centaur and in the process of transitioning to a Jupiter family comet orbit -- its orbit is relatively circular near 6 au and it is observed to be constantly active, with regular large scale outbursts \citep{Miles2016}, so it has clearly been processed by activity in some way. Naively, one might expect 29P to be less likely to retain hyper-volatiles such as CO, compared with a new LPC like E1, whose activity has presumably only started very recently as it approaches the Sun. \citet{Faggi2024} also observe significant spatial structure in the emission of CO and CO$_2$ in 29P.  Their coma model indicates three distinct regions of activity with local CO-to-CO$_2$ ratios ranging from 0 to 260, implying heterogeneity of the (near-)surface ices driving this activity. This could be explained by the evolution of 29P's nucleus during its prolonged period of violent activity in its current orbit. In this sense the lack of structure in the outgassing in E1 makes sense -- a new comet could be expected to be more uniformly active across its surface.

The lack of CO emission in the JWST spectrum suggests that CO is not readily available in the nucleus layer that is currently driving the activity of E1. This could be because CO was not incorporated in the bulk material, but it is more likely that it has been locally depleted due to alteration of superficial material. In the Oort cloud, comet nuclei are subject to irradiation by energetic particles such as cosmic rays which could produce a volatile-depleted and hardened superficial crust of several meters \citep{Stern1988,Stern2003}. Before comet nuclei are stored in the Oort Cloud, they were formed closer to the Sun where thermal processing could have led to some hyper-volatile loss. \citet{Lisse2022} predicted a complete loss of all CO not trapped in a more refractory ice in Oort Cloud comets. However they assumed that the ejection process would start after hundreds of Myr, when there is strong evidence that the giant planet instability triggering the outward ejection of planetesimals occurred quite rapidly \citep{Nesvorny2018}. \citet{Davidsson2021} suggested that CO would be completely lost for all small nuclei, but noted a prolonged timescale for CO loss in larger objects. This implies that the timing of the orbital evolution leading to the implantation of comet nuclei in the Oort Cloud with respect to the giant planet instability could play a critical role in the survival of CO. Indeed, \citet{Gkotsinas2024} confirm that Oort Cloud comets are relatively protected from stringent thermal processing as a result of a quick outward scattering after the giant planet instability, with $\sim$60\% of the Oort Cloud nuclei retaining some of their pure condensed CO ice. Overall, thermal processing sustained before E1's nucleus was ejected toward the Oort Cloud could explain why CO appears depleted now that it is returning to the inner solar system.

We note that a variety of thermal processing outcomes is nonetheless observed by \citet{Gkotsinas2024}, as it is tightly linked with the unique dynamical path followed by each nucleus on its way to the Oort Cloud. Indeed, the authors suggest that nuclei implanted beyond the Oort spike (that would typically be categorised as dynamically new comets when they return) are more processed in terms of CO content than those implanted closer to the Sun (which would more likely be labeled as returning comets even on their first journey back). At first glance, this could explain why some dynamically new comets tend to be less CO-dominated than returning comets \citep{AHearn2012,HarringtonPinto2022}, and a possible explanation for the lack of CO in E1. Monitoring the activity pattern and the emission of hypervolatiles in the coming months with our second and third JWST epochs will help understand whether CO can be mobilised from the subsurface material, which in turn will help better understand the early phases of that comet formation and evolution. Moreover, a sample of LPCs observed with JWST at different heliocentric distances would be beneficial to reduce the influence of individual comets' orbits and histories in order to test the hypotheses for different volatile compositions of different comet populations.  Finally, detailed dynamical studies of E1's orbit similar to those in the CODE database \citep{Krolikowska2020} and based on astrometric observations obtained throughout its orbit will help to define the recent orbital history and the possibility of previous perihelia warm enough for CO sublimation.

Turning to the absorption features, the 3 \micron{} water ice band, and in particular the Fresnel peak, is qualitatively similar in appearance to those of the water-dominated trans-Neptunian Objects (TNOs) and inactive Centaurs observed with JWST as part of the large sample `DiSCo' programme \citep{Pinilla-Alonso2024,Licandro2024,DePra2025}. The Fresnel peak is not seen in JWST spectra of active Centaurs like 39P \citep{HarringtonPinto2023} or 29P \citep{Faggi2024}.  The DiSCo TNOs and Centaurs are not active bodies, and are likely larger. They show surface crystalline water ice, while our IFU spatial observations show that the water ice in E1 is on coma particles. 
Neither the water-dominated TNOs/Centaurs nor E1 show CO ice absorption features, while other DiSCo targets do \citep{Pinilla-Alonso2024}, but, as noted above, surface CO is expected to be lost due to thermal evolution, so we cannot conclude that the comet or the TNOs/Centaurs lack CO in their interiors based on their icy coma particle or surface composition alone, and we may yet see CO emission from E1 as the comet gets closer to the Sun.

The Fresnel peak is characteristic of crystalline water ice and arises when a transition between volume and surface scattering occurs, possibly suggesting that the icy coma particles are large (relative to the few micron wavelength). Detailed spectral modelling for E1 is beyond the scope of this letter, but comparison with  modelling of the two water-dominated inactive Centaurs by \citet{Licandro2024} could  be informative: 
The presence of CO$_2$ ice in absorption in the Centaur spectra (the CO$_2$ fundamental band at 4.27 \micron) suggests that the surface is likely composed of about 50\% ices (water and CO$_2$), about 30\% silicates and 10 to 20\% of complex organics. CO$_2$ and water are suggested to be in intimate mixture: if this is also the case for the (sub)-surface active layer of E1, that would be consistent with the release of water and dust grains as CO$_2$ sublimates, and support the idea that Oort cloud comets and Centaurs formed in the same region \citep[e.g.][]{Nesvorny2018,Gkotsinas2024}. With the caveat that we observe the coma rather than the surface, comparison between the comet and the JWST-observed Centaurs samples the effects of different nucleus size and different orbital evolution for bodies from the same part of the disc.

Assuming the solar energy input is distributed between thermal re-radiation and sublimation of CO$_2$, we find an equilibrium temperature of $T_\mathrm{eq} = 98$ K and a CO$_2$ production rate of $4.7 \times 10^{20}$ molecules m$^{-2}$ s$^{-1}$ at 7 au. With  our measured $Q(CO_2) = 2.546 \pm 0.019 \times 10^{25}$ molecules s$^{-1}$, this implies a small active area, with radius $\sim$ 130 m, or 0.01\% of the surface if the nucleus is as large as our radius $\sim 13.7$ km limit -- this is in agreement with the relatively small $Af\rho$ measured from the ground. As the comet approaches the Sun we can expect an increase in activity levels as other areas become active and/or other species start to sublimate, although the rate of brightening varies significantly between comets \citep{Holt2024}. Deeper thermal wave propagation, seasonal effects, or removal of a mantled layer could result in a significant increase in gas activity above that predicted by extrapolation of this current production rate. 
We expect the Fresnel peak, and then the whole 3 \micron{} feature, will eventually disappear from the spectrum as water ice starts to sublimate, similar to the Centaurs and comets observed closer to the Sun \citep{Protopapa2018,Protopapa2021,Protopapa2022DPS,HarringtonPinto2023,Faggi2024}. 
 \citet{Protopapa2018} showed that the physical extent of the icy grain halo decreases with decreasing heliocentric distance. NIRSpec IFU provides the first opportunity to test this at meaningfully different heliocentric distances.
Whether or not CO or other highly volatile species will be seen, if they are present at depths below the surface that will eventually be heated, is more difficult to predict. Studying whether or not they are present, and also the spatial distribution within the coma if they are seen, will give important clues as to where they could be within the nucleus, which will constrain evolution models. These species could be released from amorphous water ice if there is any, or from within CO$_2$ ice \citep{Davidsson2021}, so spatial correlation between ices and gas will be important to investigate. This work presents results from the first of three JWST/NIRSpec IFU observations of this comet, at distances (7, 5, 3 au) bracketing the expected start of water ice sublimation, and we look forward (in detailed follow up paper(s)) to following its evolution to test these ideas.

\section*{Acknowledgements}

We thank the anonymous referee for helpful suggestions that improved this paper.
This work is based on observations made with the NASA/ESA/CSA James Webb Space Telescope. The data were obtained from the Mikulski Archive for Space Telescopes at the Space Telescope Science Institute, which is operated by the Association of Universities for Research in Astronomy, Inc., under NASA contract NAS 5-03127 for JWST. These observations are associated with JWST Director's Discretionary Program GO-06714.

CS acknowledges support from the UKSA. 
Support for this work was provided to CEH, MSPK,
and MMK by NASA through a grant from the Space Telescope
Science Institute, which is operated by the Association of
Universities for Research in Astronomy, Inc., under contract
NAS 5-26555.
RK acknowledges partial support by grant: K$\Pi$-06-H88/5 "Physical properties and chemical composition of asteroids and comets - a key to increasing our knowledge of the Solar System origin and evolution." by the Bulgarian National Science Fund.
AGL acknowledges support by CNES (mission Comet Interceptor).
TRAPPIST is funded by the Belgian Fund for Scientific Research (Fond National de la Recherche Scientifique, FNRS) under the grant PDR T.0120.21. EJ is a FRS-FNRS Research Director.
DF conducted this research at the Jet Propulsion Laboratory, California Institute of Technology, under a contract with the National Aeronautics and Space Administration (80NM0018D0004).
This work was supported by the International Space Science Institute (ISSI) in Bern, through ISSI International Team project 504 “The Life Cycle of Comets”.

\section*{Data Availability}


All JWST is available through the Mikulski Archive for Space Telescopes at the Space Telescope Science Institute, and directly via \url{https://doi.org/10.17909/7wnt-k845}.



\bibliographystyle{mnras}
\bibliography{2024E1} 

\begin{thebibliography}{}
\makeatletter
\relax
\def\mn@urlcharsother{\let\do\@makeother \do\$\do\&\do\#\do\^\do\_\do\%\do\~}
\def\mn@doi{\begingroup\mn@urlcharsother \@ifnextchar [ {\mn@doi@}
  {\mn@doi@[]}}
\def\mn@doi@[#1]#2{\def\@tempa{#1}\ifx\@tempa\@empty \href
  {http://dx.doi.org/#2} {doi:#2}\else \href {http://dx.doi.org/#2} {#1}\fi
  \endgroup}
\def\mn@eprint#1#2{\mn@eprint@#1:#2::\@nil}
\def\mn@eprint@arXiv#1{\href {http://arxiv.org/abs/#1} {{\tt arXiv:#1}}}
\def\mn@eprint@dblp#1{\href {http://dblp.uni-trier.de/rec/bibtex/#1.xml}
  {dblp:#1}}
\def\mn@eprint@#1:#2:#3:#4\@nil{\def\@tempa {#1}\def\@tempb {#2}\def\@tempc
  {#3}\ifx \@tempc \@empty \let \@tempc \@tempb \let \@tempb \@tempa \fi \ifx
  \@tempb \@empty \def\@tempb {arXiv}\fi \@ifundefined
  {mn@eprint@\@tempb}{\@tempb:\@tempc}{\expandafter \expandafter \csname
  mn@eprint@\@tempb\endcsname \expandafter{\@tempc}}}

\bibitem[\protect\citeauthoryear{{A'Hearn} et~al.,}{{A'Hearn}
  et~al.}{2012}]{AHearn2012}
{A'Hearn} M.~F.,  et~al., 2012, \mn@doi [\apj] {10.1088/0004-637X/758/1/29},
  \href {https://ui.adsabs.harvard.edu/abs/2012ApJ...758...29A} {758, 29}

\bibitem[\protect\citeauthoryear{Bodewits, Bonev, Cordiner  \&
  Villanueva}{Bodewits et~al.}{2024}]{Bodewits-CometsIII}
Bodewits D.,  Bonev B.~P.,  Cordiner M.~A.,   Villanueva G.~L.,  2024, in Meech
  K.~J.,  Combi M.~R.,  Bockelée-Morvan D.,  Raymond S.~N.,   Zolensky M.~E.,
  eds, , Comets III.
University of Arizona Press, pp 407--432 (\mn@eprint {arXiv} {2209.02616})

\bibitem[\protect\citeauthoryear{{Davidsson}}{{Davidsson}}{2021}]{Davidsson2021}
{Davidsson} B. J.~R.,  2021, \mn@doi [\mnras] {10.1093/mnras/stab1593}, \href
  {https://ui.adsabs.harvard.edu/abs/2021MNRAS.505.5654D} {505, 5654}

\bibitem[\protect\citeauthoryear{{De Pr{\'a}} et~al.,}{{De Pr{\'a}}
  et~al.}{2025}]{DePra2025}
{De Pr{\'a}} M.~N.,  et~al., 2025, \mn@doi [Nature Astronomy]
  {10.1038/s41550-024-02276-x}, \href
  {https://ui.adsabs.harvard.edu/abs/2025NatAs...9..252D} {9, 252}

\bibitem[\protect\citeauthoryear{{Faggi} et~al.,}{{Faggi}
  et~al.}{2024}]{Faggi2024}
{Faggi} S.,  et~al., 2024, \mn@doi [Nature Astronomy]
  {10.1038/s41550-024-02319-3}, \href
  {https://ui.adsabs.harvard.edu/abs/2024NatAs...8.1237F} {8, 1237}

\bibitem[\protect\citeauthoryear{{Farnham}, {Kelley}  \& {Bauer}}{{Farnham}
  et~al.}{2021}]{Farnham2021}
{Farnham} T.~L.,  {Kelley} M. S.~P.,   {Bauer} J.~M.,  2021, \mn@doi [\psj]
  {10.3847/PSJ/ac323d}, \href
  {https://ui.adsabs.harvard.edu/abs/2021PSJ.....2..236F} {2, 236}

\bibitem[\protect\citeauthoryear{{Gkotsinas}, {Nesvorn{\'y}},
  {Guilbert-Lepoutre}, {Raymond}  \& {Kaib}}{{Gkotsinas}
  et~al.}{2024}]{Gkotsinas2024}
{Gkotsinas} A.,  {Nesvorn{\'y}} D.,  {Guilbert-Lepoutre} A.,  {Raymond} S.~N.,
   {Kaib} N.,  2024, \mn@doi [\psj] {10.3847/PSJ/ad7f4e}, \href
  {https://ui.adsabs.harvard.edu/abs/2024PSJ.....5..243G} {5, 243}

\bibitem[\protect\citeauthoryear{{Harrington Pinto}, {Womack}, {Fernandez}  \&
  {Bauer}}{{Harrington Pinto} et~al.}{2022}]{HarringtonPinto2022}
{Harrington Pinto} O.,  {Womack} M.,  {Fernandez} Y.,   {Bauer} J.,  2022,
  \mn@doi [\psj] {10.3847/PSJ/ac960d}, \href
  {https://ui.adsabs.harvard.edu/abs/2022PSJ.....3..247H} {3, 247}

\bibitem[\protect\citeauthoryear{{Harrington Pinto} et~al.,}{{Harrington Pinto}
  et~al.}{2023}]{HarringtonPinto2023}
{Harrington Pinto} O.,  et~al., 2023, \mn@doi [\psj] {10.3847/PSJ/acf928},
  \href {https://ui.adsabs.harvard.edu/abs/2023PSJ.....4..208H} {4, 208}

\bibitem[\protect\citeauthoryear{{Holt} et~al.,}{{Holt}
  et~al.}{2024}]{Holt2024}
{Holt} C.~E.,  et~al., 2024, \mn@doi [\psj] {10.3847/PSJ/ad8e38}, \href
  {https://ui.adsabs.harvard.edu/abs/2024PSJ.....5..273H} {5, 273}

\bibitem[\protect\citeauthoryear{{Hsieh} et~al.,}{{Hsieh}
  et~al.}{2025}]{Hsieh2025}
{Hsieh} H.~H.,  et~al., 2025, \mn@doi [\psj] {10.3847/PSJ/ad9199}, \href
  {https://ui.adsabs.harvard.edu/abs/2025PSJ.....6....3H} {6, 3}

\bibitem[\protect\citeauthoryear{{Hui}, {Jewitt}  \& {Clark}}{{Hui}
  et~al.}{2018}]{Hui2018}
{Hui} M.-T.,  {Jewitt} D.,   {Clark} D.,  2018, \mn@doi [\aj]
  {10.3847/1538-3881/aa9be1}, \href
  {https://ui.adsabs.harvard.edu/abs/2018AJ....155...25H} {155, 25}

\bibitem[\protect\citeauthoryear{{Hui}, {Farnocchia}  \& {Micheli}}{{Hui}
  et~al.}{2019}]{Hui2019}
{Hui} M.-T.,  {Farnocchia} D.,   {Micheli} M.,  2019, \mn@doi [\aj]
  {10.3847/1538-3881/ab0e09}, \href
  {https://ui.adsabs.harvard.edu/abs/2019AJ....157..162H} {157, 162}

\bibitem[\protect\citeauthoryear{{Hui}, {Weryk}, {Micheli}, {Huang}  \&
  {Wainscoat}}{{Hui} et~al.}{2024}]{Hui2024}
{Hui} M.-T.,  {Weryk} R.,  {Micheli} M.,  {Huang} Z.,   {Wainscoat} R.,  2024,
  \mn@doi [\aj] {10.3847/1538-3881/ad2500}, \href
  {https://ui.adsabs.harvard.edu/abs/2024AJ....167..140H} {167, 140}

\bibitem[\protect\citeauthoryear{{Ivezi{\'c}} et~al.,}{{Ivezi{\'c}}
  et~al.}{2019}]{LSST}
{Ivezi{\'c}} {\v{Z}}.,  et~al., 2019, \mn@doi [\apj]
  {10.3847/1538-4357/ab042c}, \href
  {https://ui.adsabs.harvard.edu/abs/2019ApJ...873..111I} {873, 111}

\bibitem[\protect\citeauthoryear{{Jakobsen} et~al.,}{{Jakobsen}
  et~al.}{2022}]{NIRSpec}
{Jakobsen} P.,  et~al., 2022, \mn@doi [\aap] {10.1051/0004-6361/202142663},
  \href {https://ui.adsabs.harvard.edu/abs/2022A&A...661A..80J} {661, A80}

\bibitem[\protect\citeauthoryear{{Jewitt}, {Kim}, {Mutchler}, {Agarwal}, {Li}
  \& {Weaver}}{{Jewitt} et~al.}{2021}]{Jewitt2021}
{Jewitt} D.,  {Kim} Y.,  {Mutchler} M.,  {Agarwal} J.,  {Li} J.,   {Weaver} H.,
   2021, \mn@doi [\aj] {10.3847/1538-3881/abe4cf}, \href
  {https://ui.adsabs.harvard.edu/abs/2021AJ....161..188J} {161, 188}

\bibitem[\protect\citeauthoryear{{Jones} et~al.,}{{Jones}
  et~al.}{2024}]{Jones-etal-CI}
{Jones} G.~H.,  et~al., 2024, \mn@doi [\ssr] {10.1007/s11214-023-01035-0},
  \href {https://ui.adsabs.harvard.edu/abs/2024SSRv..220....9J} {220, 9}

\bibitem[\protect\citeauthoryear{{Kelley} et~al.,}{{Kelley}
  et~al.}{2016}]{Kelley2016}
{Kelley} M. S.~P.,  et~al., 2016, \mn@doi [\pasp]
  {10.1088/1538-3873/128/959/018009}, \href
  {https://ui.adsabs.harvard.edu/abs/2016PASP..128a8009K} {128, 018009}

\bibitem[\protect\citeauthoryear{{Kelley}, {Hsieh}, {Bodewits}, {Saki},
  {Villanueva}, {Milam}  \& {Hammel}}{{Kelley} et~al.}{2023}]{Kelley2023}
{Kelley} M. S.~P.,  {Hsieh} H.~H.,  {Bodewits} D.,  {Saki} M.,  {Villanueva}
  G.~L.,  {Milam} S.~N.,   {Hammel} H.~B.,  2023, \mn@doi [\nat]
  {10.1038/s41586-023-06152-y}, \href
  {https://ui.adsabs.harvard.edu/abs/2023Natur.619..720K} {619, 720}

\bibitem[\protect\citeauthoryear{{Knight}, {Kokotanekova}  \&
  {Samarasinha}}{{Knight} et~al.}{2024}]{Knight-CometsIII}
{Knight} M.~M.,  {Kokotanekova} R.,   {Samarasinha} N.~H.,  2024, in Meech
  K.~J.,  Combi M.~R.,  Bockelée-Morvan D.,  Raymond S.~N.,   Zolensky M.~E.,
  eds, , Comets III.
University of Arizona Press, pp 361--404 (\mn@eprint {arXiv} {2304.09309})

\bibitem[\protect\citeauthoryear{{Kr{\'o}likowska} \&
  {Dybczy{\'n}ski}}{{Kr{\'o}likowska} \&
  {Dybczy{\'n}ski}}{2020}]{Krolikowska2020}
{Kr{\'o}likowska} M.,  {Dybczy{\'n}ski} P.~A.,  2020, \mn@doi [\aap]
  {10.1051/0004-6361/202038451}, \href
  {https://ui.adsabs.harvard.edu/abs/2020A&A...640A..97K} {640, A97}

\bibitem[\protect\citeauthoryear{{Licandro} et~al.,}{{Licandro}
  et~al.}{2025}]{Licandro2024}
{Licandro} J.,  et~al., 2025, \mn@doi [Nature Astronomy]
  {10.1038/s41550-024-02417-2}, \href
  {https://ui.adsabs.harvard.edu/abs/2025NatAs...9..245L} {9, 245}

\bibitem[\protect\citeauthoryear{{Lisse} et~al.,}{{Lisse}
  et~al.}{2022}]{Lisse2022}
{Lisse} C.~M.,  et~al., 2022, \mn@doi [\psj] {10.3847/PSJ/ac6097}, \href
  {https://ui.adsabs.harvard.edu/abs/2022PSJ.....3..112L} {3, 112}

\bibitem[\protect\citeauthoryear{{Lister} et~al.,}{{Lister}
  et~al.}{2022}]{Lister-LOOK-summary}
{Lister} T.,  et~al., 2022, \mn@doi [\psj] {10.3847/PSJ/ac7a31}, \href
  {https://ui.adsabs.harvard.edu/abs/2022PSJ.....3..173L} {3, 173}

\bibitem[\protect\citeauthoryear{{Mastrapa}, {Sandford}, {Roush}, {Cruikshank}
  \& {Dalle Ore}}{{Mastrapa} et~al.}{2009}]{Mastrapa2009}
{Mastrapa} R.~M.,  {Sandford} S.~A.,  {Roush} T.~L.,  {Cruikshank} D.~P.,
  {Dalle Ore} C.~M.,  2009, \mn@doi [\apj] {10.1088/0004-637X/701/2/1347},
  \href {https://ui.adsabs.harvard.edu/abs/2009ApJ...701.1347M} {701, 1347}

\bibitem[\protect\citeauthoryear{{Mastrapa}, {Grundy}  \&
  {Gudipati}}{{Mastrapa} et~al.}{2013}]{Mastrapa2013}
{Mastrapa} R. M.~E.,  {Grundy} W.~M.,   {Gudipati} M.~S.,  2013, in {Gudipati}
  M.~S.,  {Castillo-Rogez} J.,  eds,  Astrophysics and Space Science Library
  Vol. 356, Astrophysics and Space Science Library. p.~371,
  \mn@doi{10.1007/978-1-4614-3076-6_11}

\bibitem[\protect\citeauthoryear{{Mazzotta Epifani}, {Palumbo}  \&
  {Colangeli}}{{Mazzotta Epifani} et~al.}{2009}]{MazzottaEpifani2009}
{Mazzotta Epifani} E.,  {Palumbo} P.,   {Colangeli} L.,  2009, \mn@doi [\aap]
  {10.1051/0004-6361/200912611}, \href
  {https://ui.adsabs.harvard.edu/abs/2009A&A...508.1031M} {508, 1031}

\bibitem[\protect\citeauthoryear{{Meech} \& {Svoren}}{{Meech} \&
  {Svoren}}{2004}]{Meech-CometsII}
{Meech} K.~J.,  {Svoren} J.,  2004, in {Festou} M.~C.,  {Keller} H.~U.,
  {Weaver} H.~A.,  eds, , Comets II.
p.~317

\bibitem[\protect\citeauthoryear{{Meech} et~al.,}{{Meech}
  et~al.}{2009}]{Meech2009}
{Meech} K.~J.,  et~al., 2009, \mn@doi [\icarus] {10.1016/j.icarus.2008.12.045},
  \href {https://ui.adsabs.harvard.edu/abs/2009Icar..201..719M} {201, 719}

\bibitem[\protect\citeauthoryear{{Miles}, {Faillace}, {Mottola}, {Raab},
  {Roche}, {Soulier}  \& {Watkins}}{{Miles} et~al.}{2016}]{Miles2016}
{Miles} R.,  {Faillace} G.~A.,  {Mottola} S.,  {Raab} H.,  {Roche} P.,
  {Soulier} J.-F.,   {Watkins} A.,  2016, \mn@doi [\icarus]
  {10.1016/j.icarus.2015.11.019}, \href
  {https://ui.adsabs.harvard.edu/abs/2016Icar..272..327M} {272, 327}

\bibitem[\protect\citeauthoryear{Nelson, Lide  \& Maryott}{Nelson
  et~al.}{1967}]{Nelson_1967}
Nelson R.~D.,  Lide D.~R.,   Maryott A.~A.,  1967, Selected values of electric
  dipole moments for molecules in the gas phase, \mn@doi{10.6028/nbs.nsrds.10.
}, \url {http://dx.doi.org/10.6028/nbs.nsrds.10}

\bibitem[\protect\citeauthoryear{{Nesvorn{\'y}}, {Vokrouhlick{\'y}}, {Bottke}
  \& {Levison}}{{Nesvorn{\'y}} et~al.}{2018}]{Nesvorny2018}
{Nesvorn{\'y}} D.,  {Vokrouhlick{\'y}} D.,  {Bottke} W.~F.,   {Levison} H.~F.,
  2018, \mn@doi [Nature Astronomy] {10.1038/s41550-018-0564-3}, \href
  {https://ui.adsabs.harvard.edu/abs/2018NatAs...2..878N} {2, 878}

\bibitem[\protect\citeauthoryear{{Pinilla-Alonso} et~al.,}{{Pinilla-Alonso}
  et~al.}{2024}]{Pinilla-Alonso2024-Chiron}
{Pinilla-Alonso} N.,  et~al., 2024, \mn@doi [\aap]
  {10.1051/0004-6361/202450124}, \href
  {https://ui.adsabs.harvard.edu/abs/2024A&A...692L..11P} {692, L11}

\bibitem[\protect\citeauthoryear{{Pinilla-Alonso} et~al.,}{{Pinilla-Alonso}
  et~al.}{2025}]{Pinilla-Alonso2024}
{Pinilla-Alonso} N.,  et~al., 2025, \mn@doi [Nature Astronomy]
  {10.1038/s41550-024-02433-2}, \href
  {https://ui.adsabs.harvard.edu/abs/2025NatAs...9..230P} {9, 230}

\bibitem[\protect\citeauthoryear{{Protopapa}, {Kelley}, {Yang}, {Bauer},
  {Kolokolova}, {Woodward}, {Keane}  \& {Sunshine}}{{Protopapa}
  et~al.}{2018}]{Protopapa2018}
{Protopapa} S.,  {Kelley} M. S.~P.,  {Yang} B.,  {Bauer} J.~M.,  {Kolokolova}
  L.,  {Woodward} C.~E.,  {Keane} J.~V.,   {Sunshine} J.~M.,  2018, \mn@doi
  [\apjl] {10.3847/2041-8213/aad33b}, \href
  {https://ui.adsabs.harvard.edu/abs/2018ApJ...862L..16P} {862, L16}

\bibitem[\protect\citeauthoryear{{Protopapa}, {Kelley}, {Woodward}  \&
  {Yang}}{{Protopapa} et~al.}{2021}]{Protopapa2021}
{Protopapa} S.,  {Kelley} M. S.~P.,  {Woodward} C.~E.,   {Yang} B.,  2021,
  \mn@doi [\psj] {10.3847/PSJ/ac135a}, \href
  {https://ui.adsabs.harvard.edu/abs/2021PSJ.....2..176P} {2, 176}

\bibitem[\protect\citeauthoryear{{Protopapa}, {Kelley}, {Yang}  \&
  {Gustafsson}}{{Protopapa} et~al.}{2022}]{Protopapa2022DPS}
{Protopapa} S.,  {Kelley} M.,  {Yang} B.,   {Gustafsson} A.,  2022, in
  AAS/Division for Planetary Sciences Meeting Abstracts. p. 109.04

\bibitem[\protect\citeauthoryear{{Rousselot}, {Korsun}, {Kulyk}, {Afanasiev},
  {Ivanova}, {Sergeev}  \& {Velichko}}{{Rousselot}
  et~al.}{2014}]{Rousselot2014}
{Rousselot} P.,  {Korsun} P.~P.,  {Kulyk} I.~V.,  {Afanasiev} V.~L.,  {Ivanova}
  O.~V.,  {Sergeev} A.~V.,   {Velichko} S.~F.,  2014, \mn@doi [\aap]
  {10.1051/0004-6361/201424223}, \href
  {https://ui.adsabs.harvard.edu/abs/2014A&A...571A..73R} {571, A73}

\bibitem[\protect\citeauthoryear{{S{\'a}nchez} et~al.,}{{S{\'a}nchez}
  et~al.}{2021}]{Pau-CI}
{S{\'a}nchez} J.~P.,  et~al., 2021, \mn@doi [Acta Astronautica]
  {10.1016/j.actaastro.2021.07.014}, \href
  {https://ui.adsabs.harvard.edu/abs/2021AcAau.188..265S} {188, 265}

\bibitem[\protect\citeauthoryear{{Shostak}, {Ebenstein}  \&
  {Muenter}}{{Shostak} et~al.}{1991}]{Shostak1991}
{Shostak} S.~L.,  {Ebenstein} W.~L.,   {Muenter} J.~S.,  1991, \mn@doi [\jcp]
  {10.1063/1.460471}, \href
  {https://ui.adsabs.harvard.edu/abs/1991JChPh..94.5875S} {94, 5875}

\bibitem[\protect\citeauthoryear{{Snodgrass} \& {Jones}}{{Snodgrass} \&
  {Jones}}{2019}]{Snodgrass+Jones-CI}
{Snodgrass} C.,  {Jones} G.~H.,  2019, \mn@doi [Nature Communications]
  {10.1038/s41467-019-13470-1}, \href
  {https://ui.adsabs.harvard.edu/abs/2019NatCo..10.5418S} {10, 5418}

\bibitem[\protect\citeauthoryear{{Stern}}{{Stern}}{2003}]{Stern2003}
{Stern} S.~A.,  2003, \nat, \href
  {https://ui.adsabs.harvard.edu/abs/2003Natur.424..639S} {424, 639}

\bibitem[\protect\citeauthoryear{{Stern} \& {Shull}}{{Stern} \&
  {Shull}}{1988}]{Stern1988}
{Stern} S.~A.,  {Shull} J.~M.,  1988, \mn@doi [\nat] {10.1038/332407a0}, \href
  {https://ui.adsabs.harvard.edu/abs/1988Natur.332..407S} {332, 407}

\bibitem[\protect\citeauthoryear{{Villanueva}, {Smith}, {Protopapa}, {Faggi}
  \& {Mandell}}{{Villanueva} et~al.}{2018}]{Villanueva-PSG}
{Villanueva} G.~L.,  {Smith} M.~D.,  {Protopapa} S.,  {Faggi} S.,   {Mandell}
  A.~M.,  2018, \mn@doi [\jqsrt] {10.1016/j.jqsrt.2018.05.023}, \href
  {https://ui.adsabs.harvard.edu/abs/2018JQSRT.217...86V} {217, 86}

\bibitem[\protect\citeauthoryear{{Wierzchos} et~al.,}{{Wierzchos}
  et~al.}{2024}]{MPEC}
{Wierzchos} K.~W.,  et~al., 2024, \mn@doi [Minor Planet Electronic Circulars]
  {10.48377/MPEC/2024-E102}, 2024-E102

\makeatother
\end{thebibliography}

\bsp	
\label{lastpage}
\end{document}